\documentclass[journal=jacsat,manuscript=article]{achemso}
\usepackage[version=3]{mhchem} %

\usepackage{amsmath} %
\usepackage{amsfonts} %
\usepackage{amsopn} %

\usepackage{mathrsfs} %
\usepackage{euscript} %
\usepackage{mathtools}

\usepackage[dvipsnames]{xcolor}
\usepackage{graphicx} %
\usepackage{algorithm,algpseudocode}
\usepackage{lineno}

\usepackage[hidelinks]{hyperref}       
\hypersetup{
    colorlinks,
    linkcolor={blue!80!black},
    citecolor={blue!80!black},
    urlcolor={black}
}
\usepackage[capitalize,nameinlink]{cleveref} 
\usepackage{comment}

\newcommand{\diff}{\,\mathrm{d}} %
\usepackage{siunitx} %

\newcommand{\keyword}[1]{\textbf{\textbf{Keywords:}} #1}

\author{Bach Do}
\altaffiliation{These authors contributed equally to this work.}
\affiliation[University of Houston]
{Department of Civil and Environmental Engineering, University of Houston, Houston, TX 77204, USA}

\author{Sina Jafari Ghalekohneh}
\altaffiliation{These authors contributed equally to this work.}
\affiliation[University of Houston]
{Department of Mechanical and Aerospace Engineering, University of Houston, Houston, TX 77204, USA}

\author{Taiwo Adebiyi}
\affiliation[University of Houston]
{Department of Civil and Environmental Engineering, University of Houston, Houston, TX 77204, USA}

\author{Bo Zhao}
\email{bzhao8@uh.edu}
\affiliation[University of Houston]
{Department of Mechanical and Aerospace Engineering, University of Houston, Houston, TX 77204, USA}

\author{Ruda Zhang}
\email{rudaz@uh.edu}
\affiliation[University of Houston]
{Department of Civil and Environmental Engineering, University of Houston, Houston, TX 77204, USA}

\title {Automated design of nonreciprocal thermal emitters via Bayesian optimization}

\begin{document}

\begin{abstract}
    Nonreciprocal thermal emitters that break Kirchhoff’s law of thermal radiation promise exciting applications for thermal and energy applications. The design of the bandwidth and angular range of the nonreciprocal effect, which directly affects the performance of nonreciprocal emitters, typically relies on physical intuition. 
    In this study, we present a general numerical approach to maximize the nonreciprocal effect. %
    We choose doped magneto-optic materials and magnetic Weyl semimetal materials as model materials and focus on pattern-free multilayer structures. The optimization randomly starts from a less effective structure and incrementally improves the broadband nonreciprocity through the combination of Bayesian optimization and reparameterization.
    Optimization results show that the proposed approach can discover structures that can achieve
    broadband nonreciprocal emission at wavelengths from 5 to 40 \si{\um} using only a fewer layers,
    significantly outperforming current state-of-the-art designs based on intuition
    in terms of both performance and simplicity.
\end{abstract}

\keyword{Nonreciprocity, Thermal emitters, Gaussian process, Bayesian optimization, Reparameterization}

\section{Introduction}
Over the past two decades, the control of thermal radiation transfer has gained significant interest due to its vital applications in spacecraft, manufacturing, thermal management, and energy conversion\cite{Greffet2002,FanS2017,Baranov2019,LiY2021}. Typically, the design and simulation of these systems assume reciprocity. According to Kirchhoff’s law\cite{Kirchhoff1978,Planck1914}, for reciprocal thermal emitters, emissivity ($\varepsilon$) and absorptivity ($\alpha$) are equal for a given direction, frequency, and polarization. This reciprocal relationship between emission and absorption represents a significant constraint on controlling the radiative heat flow\cite{ries1983complete,snyder1998thermodynamic}, and prevents radiative energy harvesting technologies from reaching their thermodynamic limits \cite{Green2012,buddhiraju2018thermodynamic,li2020thermodynamic,park2021reaching,JafariGhalekohne2022,Park2022photonics}. 

 Recent research advances \cite{ZhuL2014,FanS2017,ZhaoB2020,shayegan2023direct} have suggested that reciprocity is not a requirement of thermodynamics, and it breaks down for nonreciprocal thermal emitters which allow $\varepsilon$ and $\alpha$ to be separately controlled. Nonreciprocal thermal emission provides the pathway to record-breaking high-efficiency radiative energy harvesting techniques\cite{Green2012,Park2022letters,JafariGhalekohne2022,Park2022photonics,ZhangZ2022,YangS2024}, thermal regulation systems \cite{ZhuL2016,LiuM2023,DongJ2021,FanL2020}, and mechanical propulsion with radiative heat flow \cite{Khandekar2021,Gelbwaser-Klimovsky2021}. Importantly, to achieve the ultimate performance in these applications, especially for far-field scenarios, nonreciprocal thermal radiative properties should be achieved over a broad wavelength and angular range. Designing nonreciprocal structures with desired properties is a new challenge compared to designing reciprocal emitters, \cite{ShiY2018,YuS2023} since one needs to codesign emissivity and absorptivity simultaneously. 

Early work has demonstrated that enhanced nonreciprocal properties \cite{zhao2019violation,ZhaoB2020} could be obtained in semiconductors and semimetals at the excitation of resonances in the wavelength range where the dielectric function crosses zero (i.e., epsilon-near-zero or ENZ range). Based on this principle, current approaches to achieving broadband nonreciprocity rely on multilayered structures in which each layer exhibits strong nonreciprocity at a different ENZ wavelength. Combined, the structure can trigger a number of these resonances simultaneously and yield broadband nonreciprocity. For example, \citeauthor{LiuM2023} \cite{LiuM2023} demonstrated broadband nonreciprocal emission at wavelengths from 20 to 40 \si{\um} using a gradient-doped multilayer InAs structure, where each layer shows strong nonreciprocity at a specific resonance wavelength under an external magnetic field. Similarly, \citeauthor{ZhangZ2023} \cite{ZhangZ2023} proposed broadband nonreciprocity at wavelengths from 10 to 20 \si{\um} using a multilayer magnetic Weyl semimetal structure with a gradient chemical potential. However, since the material and the associated parameters, like carrier concentration and geometric parameters, can all be free to choose, designing high-performance structures of this kind is quite challenging and typically relies on physical intuition. As a result, the final designs are oftentimes suboptimal; for example, requiring a high number of layers that are challenging to fabricate. Additionally, current studies usually focus on optimizing the absolute value of the contrast between emissivity and absorptivity\cite{wang2023maximal}. We make sure the sign of the contrast is respected in the design algorithm since it is crucial to the direction of photon transport \cite{JafariGhalekohne2024}  and can be important in various applications where the net photon flow direction is significant \cite{JafariGhalekohne2022,zhu2016persistent}. Despite the effectiveness of multilayer structures in achieving broadband nonreciprocal thermal emission, an effective approach to optimize the performance is lacking.

Here we propose a numerical method based on Bayesian optimization (BO) \cite{Jones1998,Brochu2010,Shahriari2016,Snoek2012,Frazier2018,Garnett2023} and reparameterization to optimize thermal emitters with a large number of design parameters and high computational cost.
BO is selected because it can address the high computational demand presented by repeateds calculations of absorptivity and emissivity during the optimization process, which hinders the use of any population-based optimizers \cite{Gold2024,ShiY2018}, gradient-based optimizers \cite{Hughes2018}, or data-driven approaches requiring a large dataset \cite{YuS2023,Kudyshev2020}.
The reparameterization imposes user-defined constraints on the structures' profiles and further reduces the number of optimization parameters, thereby reinforcing the efficiency of BO which is well-suited for small to medium optimization problems \cite{Frazier2018}.
Our optimization approach is applicable to both reciprocal and nonreciprocal emitters. In this work, we focus on demonstrating its application in achieving enhanced broadband nonreciprocal radiative properties with the sign of the contrast being maintained. We aim to find structures with a much fewer number of layers and yet can achieve nonreciprocity in an even broader wavelength range than current designs. %

\begin{figure}[t]
	\centering
	\includegraphics[width=\textwidth]{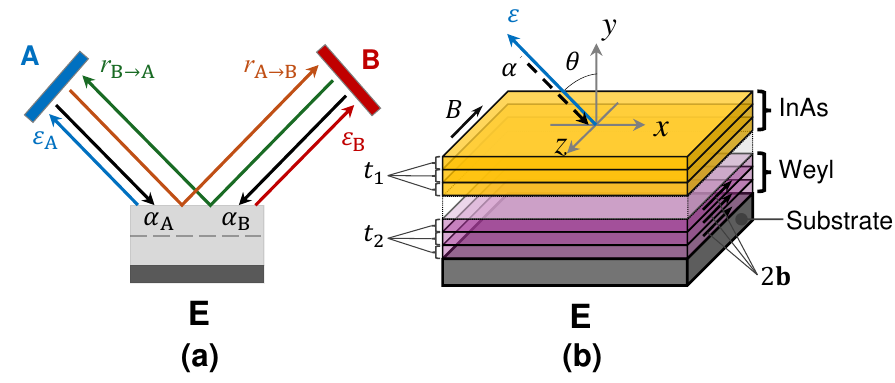}
	\caption{
 (a) Illustration of radiative heat exchange between two blackbodies (A and B) and an emitter (E) under thermal equilibrium.
 (b) Illustration of a multilayer structure consisting of InAs layers on top of Weyl semimetal layers on a reflective substrate.}
    \label{fig:EmittersStruct}
\end{figure}

\section{Absorptivity and emissivity of nonreciprocal emitters}

For opaque structures, the emissivity and absorptivity can be computed from the reflectivity, as discussed in previous studies \cite{zhu2014near,zhao2021nonreciprocal}. Here, we briefly review the rationale behind the approach to facilitate our later discussion. Consider a system in thermal equilibrium that consists of an opaque thermal emitter (E) exchanging energy with two blackbodies, A and B, (\cref{fig:EmittersStruct}(a)). The emission from blackbody A is either absorbed by the thermal emitter E ($\alpha_{\mathrm{A}}$), or reflected and absorbed by blackbody B ($r_{\mathrm{A} \rightarrow \mathrm{B}}$), leading to the following relationship:
\begin{equation} \label{eqn1}
    \alpha_{\mathrm{A}} + r_{\mathrm{A} \rightarrow \mathrm{B}} = 1,
\end{equation}
where $\alpha_{\mathrm{A}}$ represents the absorptivity in direction A at a specific wavelength, and $r_{\mathrm{A} \rightarrow \mathrm{B}}$ is the reflectivity from $\mathrm{A}$ to $\mathrm{B}$ at that wavelength.

Moreover, each blackbody absorbs and emits an equal amount of energy at thermal equilibrium, resulting in
\begin{equation} \label{eqn3}
    \begin{aligned}
        \varepsilon_{\mathrm{A}} + r_{\mathrm{B} \rightarrow \mathrm{A}} & = 1, \\
    \end{aligned}
\end{equation}

Combining Eqs.~(\ref{eqn1}) %
, and (\ref{eqn3}), we have
\begin{equation}
    \alpha_{\mathrm{A}}-\varepsilon_{\mathrm{A}} = r_{\mathrm{B} \rightarrow \mathrm{A}}-r_{\mathrm{A} \rightarrow \mathrm{B}}. %
\end{equation}

For reciprocal emitters with $r_{\mathrm{A} \rightarrow \mathrm{B}} = r_{\mathrm{B} \rightarrow \mathrm{A}}$\cite{zhao2019connection}, it follows that $\alpha_{\mathrm{A}} = \varepsilon_{\mathrm{A}}$ %
, which is consistent with Kirchhoff’s law of radiation \cite{zhang2007nano}.
However, for nonreciprocal emitters, we have $r_{\mathrm{A} \rightarrow \mathrm{B}} \neq r_{\mathrm{B} \rightarrow \mathrm{A}}$ \cite{zhu2014near}, and thus $\alpha_{\mathrm{A}} \neq \varepsilon_{\mathrm{A}}$, %
which violates Kirchhoff’s law of radiation. The absorptivity and emissivity are given as 
\begin{equation}
    \begin{aligned}
        \alpha_{\mathrm{A}} & = 1 - r_{\mathrm{A} \rightarrow \mathrm{B}},\\
        \varepsilon_{\mathrm{A}} & = 1 - r_{\mathrm{B} \rightarrow \mathrm{A}}.
    \end{aligned}
\end{equation}

For linear and static thermal emitters we consider here \cite{ghanekar2022violation,ghanekar2022nonreciprocal}, one needs to start with a material that breaks the reciprocity to obtain nonreciprocal thermal radiation. Magneto-optical material InAs\cite{zhu2014near,zhao2019violation,LiuM2023,shayegan2024broadband} and magnetic Weyl semimetals\cite{ZhaoB2020,pajovic2020intrinsic,park2021violating,picardi2024nonreciprocity} have been used for this purpose . As shown from their dielectric functions discussed later, the former works more effectively in the longer wavelength in the mid-infrared range, whereas the latter can provide nonreciprocity for shorter wavelengths. We note that, though each material has been used individually for nonreciprocal emitter design, it is not clear whether and how these materials can be combined to achieve even broader band of nonrecirpcoal thermal emission. Therefore, we choose these materials as our model nonreciprocal materials and focus on structures shown in \cref{fig:EmittersStruct}(b) that have InAs layers on top of Weyl semimetal layers on a reflective substrate, where we use silver as substrate in this study. In this way, we rule out the type of structures that possess alternating InAs and Weyl semimetal layers since they can be very challenging to fabricate in practice. We will also seek to optimize the structures containing InAs only, so that we can compare with the state-of-the-art design\cite{LiuM2023} and showcase the significant improvement through using our approach.  

We consider transverse magnetic (TM) waves with a magnetic field along the $-z$ direction, and consider the Voigt configurations\cite{zhao2019violation}. In this way, we trigger the nonreciprocity for the radiative properties at different polar angle $\theta$ in the $x$--$y$ plane. For InAs, we apply an external magnetic field ($B$ field) along the $-z$ direction \cite{Pershan1967,Aers1978}. In this case, the permittivity tensor of each InAs layer reads\cite{ZhuL2014} 
\begin{equation} \label{eqn5}
     \boldsymbol{\varepsilon} = 
     \begin{bmatrix}
 
\varepsilon_{xx} & \varepsilon_{xy} & 0 \\
\varepsilon_{yx} & \varepsilon_{yy} & 0 \\
0 & 0 & \varepsilon_{zz}

     \end{bmatrix}.
 \end{equation}
The components of the permittivity tensor for InAs can be expressed as
\begin{equation}\label{eqn6}
    \begin{aligned}
        \varepsilon_{xx} & = \varepsilon_{yy} = \varepsilon_{\infty} - \frac{\omega_p^2(\omega + i\Gamma)}{\omega\left[(\omega + i\Gamma)^2 - \omega_c^2\right]},\\
        \varepsilon_{xy} & = -\varepsilon_{yx} = \frac{i\omega_p^2\omega_c}{\omega\left[(\omega + i\Gamma)^2 - \omega_c^2\right]}, \\
        \varepsilon_{zz} & = \varepsilon_{\infty} - \frac{\omega_p^2}{\omega(\omega + i\Gamma)},
    \end{aligned}
\end{equation}
where $\omega$ is the angular frequency, $\varepsilon_\infty $ is the high-frequency permittivity, $\Gamma$ is the damping rate, $\omega_p = \sqrt{n_e e^2 / (m^* \varepsilon_0)}$ is the plasma frequency, and $\omega_c = eB / m^*$ is the cyclotron frequency. Here, $n_e$ is the carrier concentration, $e$ is the elementary charge, $m^*$ is the effective electron mass, and $\varepsilon_0$ is the vacuum permittivity. We focus on the ENZ point where the real part of ${\varepsilon_{xx}}$ crosses zero\cite{ZhuL2014,Halterman2018,Campione2015,Kinsey2019,XuJ2021}  and brings a significant enhancement to $\varepsilon_{xy}/\text{Re}(\varepsilon_{xx})$ and the nonreciprocal effect\cite{zhao2019violation}. By adjusting the carrier concentration $n_e$, which we can experimentally control while fabricating InAs among other parameters in \cref{eqn6}, the ENZ point shifts, allowing us to control the wavelength at which the ENZ region occurs.

The nonreciprocal effect in magnetic Weyl semimetal layers is intrinsic and does not require an external magnetic field\cite{ZhaoB2020}. The momentum separation of the Weyl cones, $2\mathbf{b}$, acts similarly to an applied magnetic field in magneto-optical systems. We set $\mathbf{b}=-b {\bf k}_z$ to be also along the $-z$ direction (\cref{fig:EmittersStruct}(b)) following the Voigt configuration and, in this case, the permittivity tensor of Weyl semimetal has the same format as \cref{eqn5} \cite{ZhaoB2020}.
Similar to InAs, the nonreciprocal effect for Weyl is significantly enhanced when the diagonal element $\boldsymbol{\varepsilon}$ crosses zero at ENZ point. 

The diagonal elements of the permittivity of each Weyl semimetal layer read $\varepsilon_{xx} =\varepsilon_{yy} =\varepsilon_{zz} = \varepsilon_b + \frac{i \sigma}{\omega \varepsilon_0}$, where $\varepsilon_b$ is the background permittivity, $\omega$ is the radiation frequency, and $\sigma$ is the bulk conductivity obtained from\cite{ZhaoB2020} 
\begin{equation} \label{Eqn7}
\sigma = \frac{\varepsilon_0 r_s g E_F}{6 \hbar} \Omega G\left(\frac{E_F \Omega}{2}\right) 
+ i\frac{\varepsilon_0 r_s g E_F}{6 \pi \hbar} \left\{ \frac{4}{\Omega} \left[ 1 + \frac{\pi^2}{3} \left(\frac{k_B T}{E_F}\right)^2 \right] 
+ 8 \Omega \int_{0}^{\xi_c} \frac{G(E_F \xi) - G\left(\frac{E_F \Omega}{2}\right)}{\Omega^2 - 4\xi^2} \xi d\xi\right\}.
\end{equation}
Here, $r_s = e^2 / (4\pi \varepsilon_0 \hbar \nu_F)$ is the effective fine structure constant, $\hbar$ is the reduced Planck constant, and $\nu_F$ is the Fermi velocity. $\Omega = \hbar(\omega + i \tau^{-1}) / E_F$ is the normalized complex frequency, $\tau^{-1}$ is the scattering rate, $E_F$ is the Fermi level, and $T$ is the temperature, which we assume is 300 K for this study.
$G(E) = n(-E) - n(E)$, where $n(E)$ is the Fermi-Dirac distribution function, $g$ is the number of Weyl points, and $k_B$ is the Boltzmann constant.
$\xi_c = E_C / E_F$ is the normalized cutoff energy, where $E_C$ is the cutoff energy beyond which the band dispersion is no longer linear. 
In addition, the off-diagonal element of the permittivity tensor of each Weyl semimetal layer is calculated as 
\begin{equation}
\varepsilon_{xy} = -\varepsilon_{yx} = i\frac{be^2}{2\pi^2 \hbar \omega}. 
\end{equation}
By tuning the Fermi level $E_F$—a parameter similar to $n_e$ for InAs that can be adjusted in experiments—the ENZ point 
of the Weyl semimetal can be modified, which shifts the wavelength at which the associated resonance occurs.

\begin{figure*}[t]
  \centering
  \includegraphics[scale=0.524]{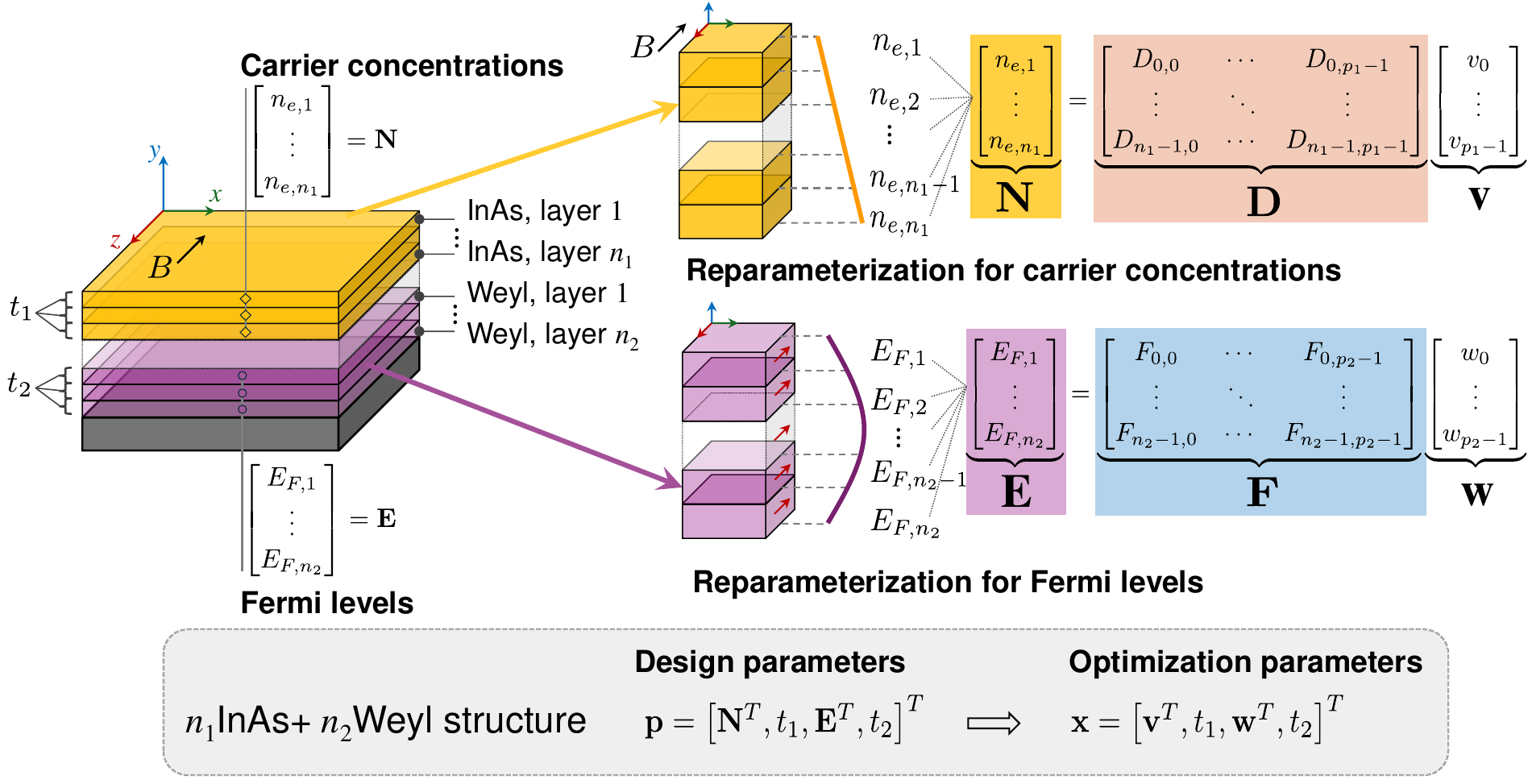}
  \caption{Design parameters ${\bf p}$ for a nonreciprocal emitter of interest and their reparameterization into ${\bf x}$.
    The structure consists of $n_1$ InAs layers with carrier concentrations ${\bf N}$ and thickness $t_1$,
    and $n_2$ Weyl semimetal layers with Fermi levels ${\bf E}$ and thickness $t_2$.
    After reparameterizing ${\bf N}$ and ${\bf E}$,
    the number of parameters reduces from $n_1+n_2+2$ to $p_1+p_2+2$,
    where $p_1 \le n_1$ and $p_2 \le n_2$ are the numbers of columns of ${\bf D}$ and ${\bf F}$, respectively.
  }
  \label{fig:Reparameterization}
\end{figure*}

\section{Optimization problem}

For the structure shown in \cref{fig:EmittersStruct}(b), each layer exhibits a distinct ENZ  point, resulting in enhanced nonreciprocity across different wavelengths. Due to the multilayer design, with each layer having a unique ENZ point, broadband nonreciprocity can be achieved over a wide wavelength range. To optimize the contrast between absorptivity and emissivity across this broad bandwidth, it is necessary to solve an optimization problem to determine the optimal values of design parameters:
the carrier concentrations for InAs layers, the Fermi levels for Weyl semimetal layers,
and the thickness of each layer in the multilayer structure. In this study, we only try to optimize the mentioned design parameters, while assuming other parameters of the dielectric functions of InAs and Weyl semimetals are fixed. For the InAs, the high-frequency permittivity and the effective electron mass of InAs are $\varepsilon_\infty = 12.37$ and $m^\star = 0.033 m_e$ \cite{madelung2004semiconductors}, respectively, where $m_e$ is the electron mass. The damping rate of InAs is set as $\Gamma = 5.9 \times 10^{12}$ rad/s \cite{LiuM2023}. The background permittivity, the number of Weyl points,  the Fermi velocity, the relaxation time, and the cutoff energy of Weyl semimetals are  $\varepsilon_b = 6.2$, $g=2$, $\tau = 10^{-12}$ s, $\nu_F = 0.83 \times 10^5$ m/s, and $E_C = 0.45$ eV, respectively \cite{ZhaoB2020}. %

Our optimization starts by defining an objective function.
Denote emitter design parameters ${\bf p} \in \mathbb{R}_+^p$,
wavelength $\lambda \in \mathbb{R}_+$, and angle of incidence $\theta \in \mathbb{R}_+$.
Let $\alpha({\bf p},\lambda,\theta)$ and $\varepsilon({\bf p},\lambda,\theta)$ 
be the local functions for the absorptivity and emissivity of this emitter, respectively.
The local function for the contrast between the absorptivity and emissivity is defined as:
\begin{equation}\label{eqn:localcontrast}
    \eta({\bf p},\lambda,\theta) = \alpha({\bf p},\lambda,\theta)-\varepsilon({\bf p},\lambda,\theta),
\end{equation}
where $\eta \in$ [-1,1]. Achieving maximum or minimum contrast is equivalent since the sign of the contrast can be flipped by switching the magnetic field. 
\Cref{fig:Reparameterization} shows the design parameters ${\bf p}$ for a nonreciprocal thermal emitter that is %
the combination of $n_1$ doped InAs layers on top of $n_2$ Weyl semimetal layers.
For this structure, ${\bf p}$ has $n_1$ distinct carrier concentrations $n_{e,i}$ $(i=1,\dots,n_1)$ for the InAs layers, the thickness $t_1$ of each InAs layer, $n_2$ distinct Fermi levels $E_{F,j}$ $(j=1,\dots,n_2)$ for the Weyl semimetal layers, and the thickness $t_2$ of each Weyl semimetal layer.
Here we simplify the optimization problem by using the same thickness for layers of each type of material.

Maintaining the sign of the contrast, we formulate the following optimization problem for the nonreciprocal emitter, aimed at finding an optimal structure that achieves negative contrast over broad wavelength and angular ranges:
\begin{equation} \label{eq:oriopt}
    \underset{ \bf p \in \mathcal{P}}{\textrm{minimize}} \quad  \widetilde{\eta}({\bf p}),
\end{equation} 
 where $\widetilde{\eta}$ represents the normalized contrast, which serves as the figure of merit for the broadband nonreciprocity, and $\mathcal{P} \subset \mathbb{R}_+^p$ is the feasible domain of ${\bf p}$.
The normalized contrast $\widetilde{\eta} \in [-1,1]$ represents the contrast averaged across the considered ranges of wavelengths and angles.
$\widetilde{\eta}$ is defined as follows:
\begin{equation}
    \widetilde{\eta}({\bf p}) = \frac{\int_{\lambda_\text{L}}^{\lambda_\text{U}} \int_{0}^{\pi/2} \eta({\bf p},\lambda,\theta) \sin{\theta} \cos{\theta} \diff\theta \diff\lambda}
    {\int_{\lambda_\text{L}}^{\lambda_\text{U}} \int_{0}^{\pi/2} \sin{\theta} \cos{\theta}  \diff\theta \diff\lambda},
\end{equation} 
where $\lambda_\text{L}$ and $\lambda_\text{U}$ are the lower and upper values of the considered wavelength range, respectively.
A numerical quadrature method for efficiently computing the integrals in the numerator of $\widetilde{\eta}({\bf p})$ is given in the Supporting Information.

We further apply a reparameterization strategy to the objective function $\widetilde{\eta}({\bf p})$ to enforce desired design constraints on the emitter structure and/or other constraints on the optimization problem.
The design constraints allow for experimental control over the profiles (e.g., linear or nonlinear) of carrier concentrations and Fermi levels inside the multilayer structures which profoundly affect the contrast value \cite{DuC2023}. 
Meanwhile, the reparameterization ensures that the number of optimization parameters remains manageable for the optimizer,
which is crucial when there are many design parameters.

The reparameterization strategy transforms $\widetilde{\eta}({\bf p})$ over the design parameter space $\bf p \in \mathcal{P}$ into $\overline{\eta}({\bf x})$ over the space of optimization parameters ${\bf x} = \phi({\bf p}) \in \mathbb{R}^d$, with the reparameterization map $\phi(\cdot): \mathbb{R}_+^p \mapsto \mathbb{R}^d$ encapsulating all desired constraints.
As a result, the optimization problem after reparameterization reads
\begin{equation}\label{eq:reparaopt}
    \underset{ {\bf x} \in \mathcal{X}}{\textrm{minimize}} \quad  \overline{\eta}({\bf x}),
\end{equation} 
where $\mathcal{X} = \{{\bf x} \in \mathbb{R}^d|\phi^{-1}({\bf x}) \in \mathcal{P}\}$
and $\overline{\eta}({\bf x}) = \widetilde{\eta}(\phi^{-1}({\bf x}))$.
Thus, instead of solving problem (\ref{eq:oriopt}) directly, we solve problem (\ref{eq:reparaopt}) and recover an optimal set of ${\bf p}$ from the resulting optimal set of ${\bf x}$.

Given a specific value of the optimization parameters ${\bf x}$ and the reparameterization map $\phi(\cdot)$, computing $\overline{\eta}({\bf x})$ is straightforward. We first recover the design parameter ${\bf p} = \phi^{-1}({\bf x})$. We then compute $\widetilde{\eta}({\bf p})$ and set $\overline{\eta}({\bf x})$ as $\widetilde{\eta}({\bf p})$.

\Cref{fig:Reparameterization} details the reparameterization technique used in this study.
This technique transforms the carrier concentrations
${\bf N} = \left[n_{e,1},\dots,n_{e,n_1} \right]^T \in \mathbb{R}^{n_1}$ for the InAs layers and Fermi levels
${\bf E} = \left[E_{F,1},\dots,E_{F,n_2} \right]^T \in \mathbb{R}^{n_2}$ for the Weyl semimetal layers
into optimization parameters ${\bf v}\in \mathbb{R}^{p_1}$ and ${\bf w} \in \mathbb{R}^{p_2}$, respectively,
using two linear maps ${\bf D} \in \mathbb{R}^{n_1 \times p_1}$ and ${\bf F} \in \mathbb{R}^{n_2 \times p_2}$
where $p_1 \le n_1$ and $p_2 \le n_2$.
By carefully designing the span of ${\bf D}$ and ${\bf F}$,
we can reduce the number of parameters for our optimization problem, which is now formulated for ${\bf v}$ and ${\bf w}$.
This reduction is crucial when $n_1$ or $n_2$ is large,
as the performance of the optimization algorithm strongly depends on the problem's input dimension.
Bayesian optimization, for example, is most effective for problems of less than 20 dimensions \cite{Frazier2018}.
By further designing the entries of each column of ${\bf D}$ or ${\bf F}$, we are able to impose specific geometrical properties on the emitter, for example, linear profiles of carrier concentrations or quadratic profiles of Fermi levels (see \cref{fig:Reparameterization}).
The detailed computations for ${\bf D}$ and ${\bf F}$ are provided in the Supporting Information.

\begin{figure*}[ht!]
\centering
\includegraphics[scale=1.1]{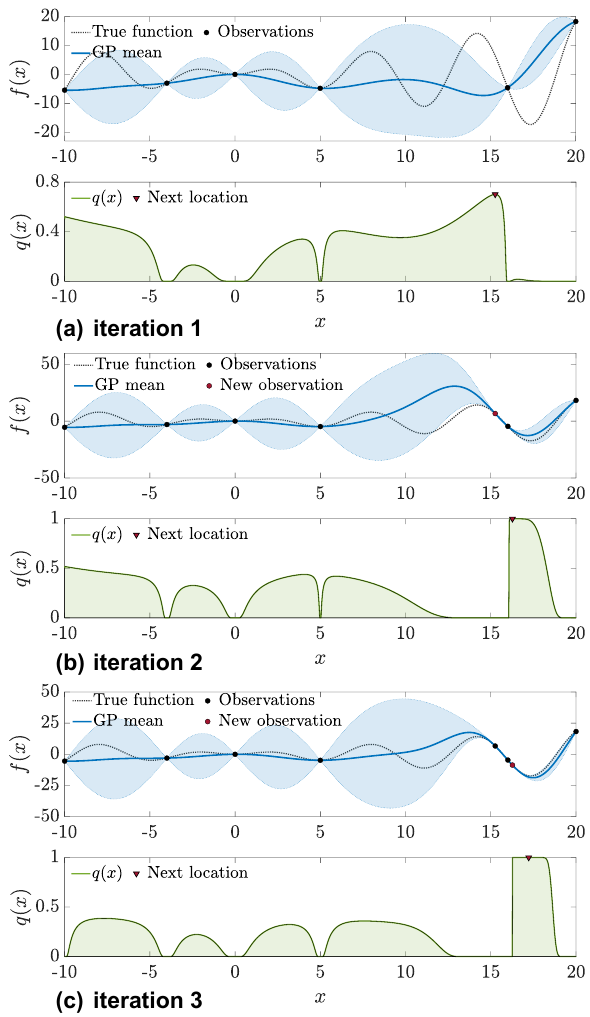}
\caption{Schematic illustration of three consecutive iterations of BO for optimizing a univariate function $f(x)$.
In each iteration, BO constructs a probabilistic model for the objective function $f(x)$ using the current data,
formulates an acquisition function $q(x)$ from the model,
and maximizes $q(x)$ to identify a new location to query the objective function.
}
\label{fig:BO}
\end{figure*}

\section{Optimizer}
Bayesian optimization (BO) \cite{Jones1998,Brochu2010,Shahriari2016,Snoek2012,Frazier2018,Garnett2023} is a successful approach to optimizing functions that are costly to evaluate, cannot be evaluated exactly, have no analytical expressions, or offer no efficient way to compute their derivatives \cite{Garnett2023}.
It finds applications in diverse domains of science and engineering, including photonic curing processes \cite{XuW2023}, adaptive experimental design \cite{Anjana2018,Stewart2020}, accelerator physics \cite{Roussel2024}, and material design \cite{Frazier2016,Vangelatos2021,Khatamsaz2023}, to name a few.
BO also enables the integration of physical and mathematical insights into the optimization process \cite{Raul2022,Do2023,ZhangRD2024mfml}.

In this study, we use BO to solve problem (\ref{eq:reparaopt}).
This is justified because the objective function $\overline{\eta}({\bf x})$ lacks a useful analytical expression and is computationally expensive.
In the following, we provide an overview of BO and explain how it can be used to incrementally improve the broadband nonreciprocity of nonreciprocal emitters starting from ineffective ones.

Given a few observations for a specific optimization problem, BO solves the problem by repeatedly
(1) constructing a probabilistic model for the objective function to represent our prior knowledge and the observations,
(2) formulating an acquisition function to define what we value in the current dataset given the probabilistic model,
and (3) maximizing the acquisition function to guide the optimization process.
This iterative process typically terminates as it reaches a prespecified number of observations we can afford,
which reflects our computational budget.
\Cref{fig:BO} illustrates three consecutive iterations of BO for minimizing a univariate objective function $f(x)$.
BO starts with six observations of $f(x)$ and locates the true global minimizer of $f(x)$ after only three iterations
(lower plot of \cref{fig:BO}c).

A Gaussian process (GP) \cite{Rasmussen2006} model often serves as the probabilistic model for BO due to its tractability and flexibility.
Typically, the GP prior is specified simply by a zero mean function and a covariance function with a closed form and a few hyperparameters.
Given a dataset of the objective function observations, the GP posterior can be derived by conditioning the GP prior on the dataset.
This GP posterior then serves as the probabilistic model that represents our beliefs about the structure of the costly objective function.
The reader is referred to the Supporting Information for the mathematical foundation of GP and the analytical formula of the GP posterior.

Once the GP posterior model has been constructed,
we formulate an acquisition function to define an optimization policy representing what we wish to exploit from this model.
The acquisition function maps each point in the design variable space to
a score on its ability to benefit the optimization process,
considering our imperfect knowledge of the objective function \cite{Jones1998,Jones2001}.
By maximizing this acquisition function, we thus maximize the potential to obtain a better new design.
We do so numerically by using, for example, a global optimization algorithm,
and the resulting solution is used in the next iteration of BO for a new evaluation of the objective function.
Maximizing the acquisition function is typically simpler and less computationally expensive than optimizing the costly objective function directly.

In \Cref{alg:BO}, we outline the use of BO for optimizing a nonreciprocal thermal emitter of interest. We begin with specifying the design parameter domain $\mathcal{P}$, the number $N$ of initial observations of the normalized contrast $\overline{\eta}({\bf x})$, and the threshold $K$ for the number of BO iterations, followed by defining the reparameterization map $\phi(\cdot)$ (\cref{alg:BO-Line1}).
To construct an initial dataset $\mathcal{D}_0$ (\cref{alg:BO-Line6}), which is necessary for initializing BO, we randomly generate a set of $N$ samples for the optimization parameters ${\bf x}$ using Latin hypercube sampling \cite{Owen1992} (\cref{alg:BO-Line2}), and subsequently evaluate $\overline{\eta}({\bf x})$ at the generated samples using rigorous coupled-wave analyses \cite{Moharam1981,yang2022mid} (\cref{alg:BO-Line4}).
Here, we assume that the evaluated values of $\overline{\eta}({\bf x})$ are noiseless, which means the numerical results perfectly capture the distribution of nonreciprocal thermal radiation in the emitter without introducing any numerical errors.

\begin{algorithm}[t]
	\caption{Bayesian optimization with noiseless observations}
	\label{alg:BO}
	\begin{algorithmic}[1]
		
		\State \textbf{input:} domain $\mathcal{P}$ of design parameters ${\bf p}$, number of initial observations $N$, threshold for the number of BO iterations $K$, reparameterization map $\phi(\cdot)$ \label{alg:BO-Line1}
		
		\State Generate $N$ initial observations of optimization parameters ${\bf x}$ \label{alg:BO-Line2}
		
		\For {$i=1:N$}  \Comment{Generate initial observations of the normalized contrast}
		\State $y_i \gets \overline{\eta}({\bf x}_i)$ \label{alg:BO-Line4}
		\EndFor
		
		\State $\mathcal{D}_0 \gets \{{\bf x}_i,y_i\}_{i=1}^N$ \Comment{Dataset of initial observations} \label{alg:BO-Line6}
		
		\State $\{{\bf x}_{\min},y_{\min}\} \gets \min\{y_i,\, i=1,\dots, N\}$ \Comment{The best observation found so far}
		
		\For {$i=N+1:N+K$}  \label{alg:BO-Line8}
            \State $k \gets i-N$
		\State Build a GP posterior $\widehat{\eta}_k({\bf x})|\mathcal{D}_{k-1}$ \label{alg:BO-Line10}
		\State Formulate an acquisition function $q({\bf x}|\mathcal{D}_{k-1})$ from $\widehat{\eta}_k({\bf x})|\mathcal{D}_{k-1}$  \label{alg:BO-Line11}
		\State ${\bf x}_{k} \gets \underset{{\bf x}}{\mathrm{arg\,max}} \ \ q({\bf x}|\mathcal{D}_{k-1})$ s.t. $\phi^{-1}({\bf x}) \in \mathcal{P}$; ${\bf x} \notin \mathcal{D}_{k-1}$ \Comment{Maximize the acquisition function} \label{alg:BO-Line12}
		\State $y_{k} \gets \overline{\eta}({\bf x}_{k})$ \Comment{Obtain a new observation of the normalized contrast} \label{alg:BO-Line13}
		\State $\mathcal{D}_k\gets\mathcal{D}_{k-1} \cup \{{\bf x}_{k},y_{k}\}$ \Comment{Update the dataset of observations} \label{alg:BO-Line14}
		\State $\{{\bf x}_{\min},y_{\min}\} \gets \min\{y_{\min},y_{k}\}$ \label{alg:BO-Line15}
        \Comment{The best observation found so far}
		\EndFor \label{alg:BO-Line16}
		
		\State \textbf{return} $\{{\bf x}_{\min},y_{\min}\}$ and ${\bf p}_{\min} \gets \phi({\bf x}_{\min})$ \label{alg:BO-Line17}
	\end{algorithmic}
\end{algorithm}

The for-loop of BO (\cref{alg:BO-Line8} to \cref{alg:BO-Line16}) commences with constructing a GP posterior $\widehat{\eta}_k({\bf x})$ for $\overline{\eta}({\bf x})$ from the current dataset $\mathcal{D}_{k-1}$ (\cref{alg:BO-Line10}), where $k=1,\dots,K$ represents the index variable for the loop. It then formulates an acquisition function $q({\bf x})$ based on $\widehat{\eta}_k({\bf x})$ (\cref{alg:BO-Line11}) and maximizes $q({\bf x})$ for a new observation ${\bf x}_k$ of the optimization parameters (\cref{alg:BO-Line12}).
To avoid reselecting points already seen in the current dataset $\mathcal{D}_{k-1}$, the constraint ${\bf x} \notin \mathcal{D}_{k-1}$ is imposed to the maximization of $q({\bf x})$ (\cref{alg:BO-Line12}).
Finally, the algorithm evaluates the normalized contrast $y_k = \overline{\eta}({\bf x}_k)$ at the new point ${\bf x}_k$, and updates the dataset with ${\bf x}_k$ and $y_k$ for use in the next iteration (\cref{alg:BO-Line13,alg:BO-Line14}).
The final optimal solution is the best observation found among points of the dataset recommended by the algorithm (\cref{alg:BO-Line15,alg:BO-Line17}).

\section{Results and discussion}

We present the optimization results for two multilayer structures of nonreciprocal thermal emitters: a 3-layer InAs structure and a 6-layer 3InAs+3Weyl structure (i.e., three layers of InAs on top of three layers of Weyl semimetal).
By considering these structures, we aim to investigate how a combination of InAs and Weyl semimetal can improve broadband nonreciprocity.  
For this purpose, we also optimize other two structures, namely a 5-layer InAs structure and an 8-layer 5InAs+3Weyl structure.
The optimization results for these structures are provided in the Supporting Information.
To evaluate the performance of the obtained optimal structures, we compare the contrast between absorptivity and emissivity of them with that of the state-of-the-art 10-layer InAs structure proposed by \citeauthor{LiuM2023} \cite{LiuM2023}

\subsection{Initialization}

We consider wavelengths $\lambda$ ranging from 5 to 40 \si{\um}, which covers the thermal infrared range and is much broader than that in existing works.
We analyze TM waves under an applied magnetic field of $B = 1.5$ T. %
The reparameterization is carefully designed for each structure to ensure that the carrier concentrations of InAs and the Weyl Fermi levels across the layers conform to a polynomial of up to second order.
As a result, there are four optimization parameters for 3-layer and 5-layer InAs structures, and eight parameters for 6-layer 3InAs+3Weyl and 8-layer 5InAs+3Weyl structures.
For 5-layer InAs and 8-layer 5InAs+3Weyl structures, the number of optimization parameters is less than the number of design parameters. 
Further details on the reparameterization scheme are provided in the Supporting Information.

There are several settings for BO and the optimizer used for maximizing the acquisition function at each BO iteration.
The GP prior is determined by a zero-mean function and the squared exponential covariance function (see Eq.~(S2) of the Supporting Information).
The carrier concentrations of InAs layers, the Fermi levels of Weyl semimetal layers, and the layer thicknesses are selected such that $n_{e,i} \in [1,10] \times 10^{17}$ atoms/$\text{cm}^3$, $E_{F,j} \in [10^{-3},0.5]$ eV, and $t_1,t_2 \in [100,5000]$ nm. The number of initial observations of the normalized contrast is $N = 5d$, where $d$ is the number of optimization parameters ${\bf x}$. The threshold for the number of BO iterations is set as $K=150$ for 3-layer InAs structure, and $K=1000$ for 6-layer 3InAs+3Weyl structure. 
BO formulates two canonical improvement-based acquisition functions: lower confidence bound (LCB) and probability of improvement (PI). Their analytical formulas are provided in the Supporting Information. To maximize each of these acquisition functions, BO uses a multi-start local optimization algorithm with 500 random starting points over the space of ${\bf x}$. The tolerance for the first-order optimality measure and the upper bound on the magnitude of any constraint functions are set at $10^{-16}$.
BO is implemented in MATLAB and executed on the Carya Cluster at the University of Houston.

\begin{figure*}[t]
	\centering
	\includegraphics[scale=0.9]{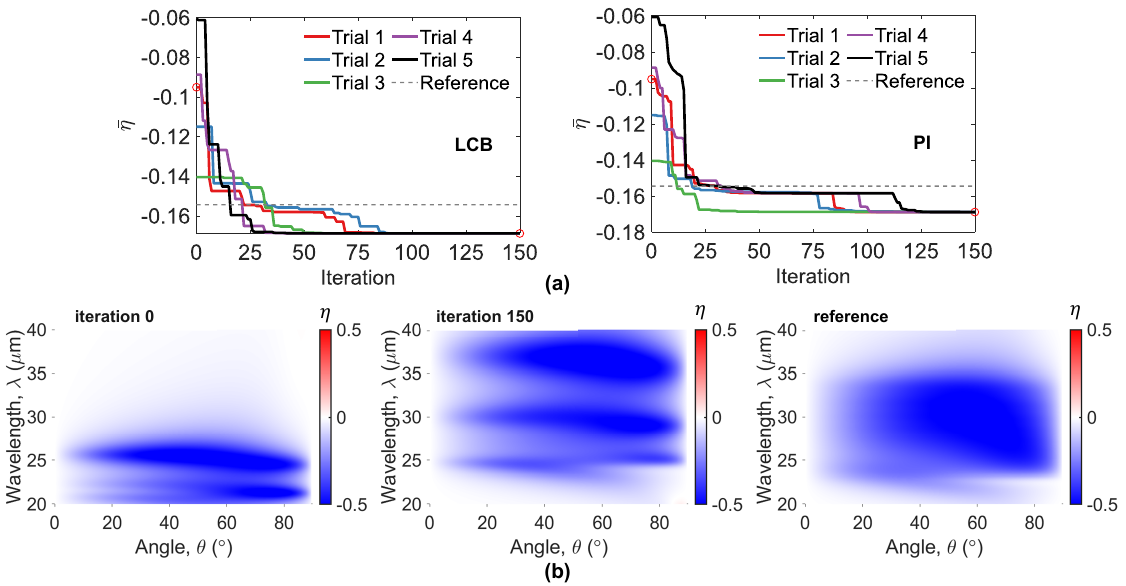}
	\caption{Optimization results for 3-layer InAs structure in comparison with the state-of-the-art 10-layer InAs structure (reference) \cite{LiuM2023}. (a) Optimization histories from LCB and PI. (b) Comparison of contrast values of the initial and final structures from the first and last iterations of the first LCB trial, and the state-of-the-art 10-layer InAs structure.}
    \label{fig:Opt3InAs}
\end{figure*}

\subsection{Optimized 3-layer InAs structure}

\Cref{fig:Opt3InAs} show the optimization results for the 3-layer InAs structure obtained from five different BO trials of LCB and five different trials of PI, with each set of trials utilizing distinct initial datasets. 
Despite starting from different initial values, all BO trials converge to a unique normalized contrast value of $\overline{\eta}=-16.9\%$ after 100 iterations (i.e., $20+100$ objective function calls) for LCB and 125 iterations (i.e., $20+125$ objective function calls) for PI, see \cref{fig:Opt3InAs}(a). 
\Cref{fig:Opt3InAs}(b) compares the contrast values of 3-layer InAs structures associated with the first and last iterations of the first BO trial of LCB with that from the state-of-the-art 10-layer InAs structure. We see that, although BO starts from a less effective initial structure (left panel), it can provide an optimal structure (middle panel) with negative contrast better than that of the state-of-the-art structure (right panel) with $\overline{\eta}=-15.4\%$, confirming its crucial role in enhancing the nonreciprocity of the 3-layer InAs structure. Moreover, it shows we can achieve substantial nonreciprocal effects with only a few layers.  

\begin{table}[t]
    \centering
    \begin{tabular}{lcc}
        Parameter & LCB & PI\\
        \hline\noalign{\smallskip}
        $n_{e,1}$ ($\times 10^{17}$ atoms/$\text{cm}^3$) & $3.353$ & $3.367$\\
        $n_{e,2}$ ($\times 10^{17}$ atoms/$\text{cm}^3$) & $5.212$ &  $5.181$\\
        $n_{e,3}$ ($\times 10^{17}$ atoms/$\text{cm}^3$) & $7.353$ & $7.367$\\
        $t_1$ (nm) & $1243$ & $1247$\\
       \hline\noalign{\smallskip}
    \end{tabular}
    \caption{Carrier concentrations of InAs and layer thickness for the best 3-layer InAs structures from LCB and PI.}
    \label{table1}
\end{table}

\cref{table1} lists the optimal parameters for the best 3-layer InAs structures obtained from LCB and PI. The optimal structures from these acquisition functions are almost identical.
Notably, the carrier concentrations of these optimal structures exhibit a linear increase from the top to the bottom layers.

\begin{figure*}[t]
	\centering
	\includegraphics[scale=1]{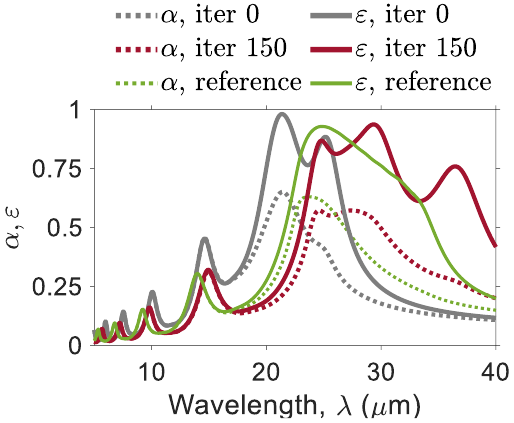}
	\caption{Absorptivity ($\alpha$) and emissivity ($\varepsilon$) of the initial and final 3-layer InAs structures from the first and last iterations of the first LCB trial, and the state-of-the-art 10-layer InAs structure (reference) \cite{LiuM2023}.}
    \label{fig:AbsorpEmit3InAs}
\end{figure*}

\Cref{fig:AbsorpEmit3InAs} shows the absorptivity and emissivity over the considered range of incidence angles for the 3-layer InAs structures from the first and last iterations of the first BO trial using LCB, as well as those of the state-of-the-art structure. We see that BO enhances the nonreciprocity of the 3-layer InAs structure by increasing the absorptivity at wavelength values from 25 to 40 \si{\um}, which correspond to the upper region of the considered spectrum.

\begin{figure*}[t]
	\centering
	\includegraphics[scale=0.9]{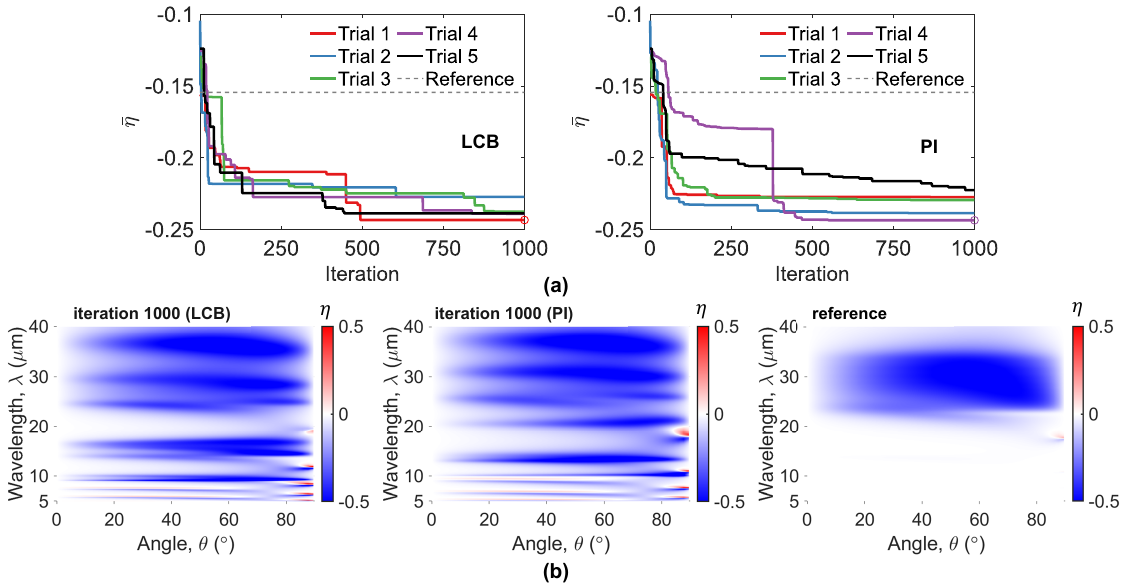}
	\caption{Optimization results for 6-layer 3InAs+3Weyl structure in comparison with the state-of-the-art 10-layer InAs structure (reference) \cite{LiuM2023}. (a) Optimization histories from LCB and PI. (b) Comparison of contrast values of the best structures from LCB and PI, and the state-of-the-art 10-layer InAs structure.}
    \label{fig:Opt3InAs3Weyl}
\end{figure*}

\begin{figure*}[t]
	\centering
	\includegraphics[scale=1]{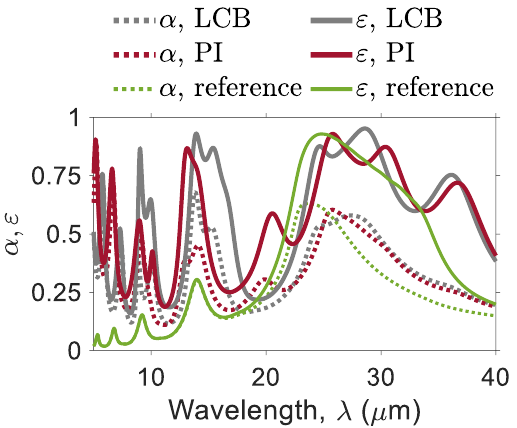}
	\caption{Absorptivity ($\alpha$) and emissivity ($\varepsilon$) of the best 6-layer 3InAs+3Weyl structures from LCB and PI, and the state-of-the-art 10-layer InAs structure (reference) \cite{LiuM2023}.}
    \label{fig:AbsorpEmit3InAs3Weyl}
\end{figure*}

\begin{table}[t]
    \centering
    \begin{tabular}{lcc}
        Parameter & LCB & PI\\
        \hline\noalign{\smallskip}
        $n_{e,1}$ ($\times 10^{17}$ atoms/$\text{cm}^3$) & $3.416$ & $3.308$\\
        $n_{e,2}$ ($\times 10^{17}$ atoms/$\text{cm}^3$) & $5.454$ &  $4.861$\\
        $n_{e,3}$ ($\times 10^{17}$ atoms/$\text{cm}^3$) & $7.416$ & $6.777$\\
        $t_1$ (nm) & $1203$ & $1066$\\
        $E_{F,1}$ (eV) & $0.097$ & $0.061$\\
        $E_{F,2}$ (eV) & $0.148$ & $0.103$\\
        $E_{F,3}$ (eV) & $0.097$ & $0.132$\\
        $t_2$ (nm) & $100$ & $100$\\
       \hline\noalign{\smallskip}
    \end{tabular}
    \caption{Carrier concentrations of InAs, Weyl Fermi levels, and layer thicknesses for the best 6-layer 3InAs+3Weyl structures from LCB and PI.}
    \label{table2}
\end{table}

\subsection{Optimized 6-layer 3InAs+3Weyl structure}

\Cref{fig:Opt3InAs3Weyl} presents the optimization results for the 6-layer 3InAs+3Weyl structure.
Although different BO trials for this structure using LCB and PI do not converge to a single contrast value after 1000 iterations, they demonstrate significant improvements in the nonreciprocity.
The best structures recommended by LCB and PI achieve the same negative contrast value of $\overline{\eta} = -24.4\%$, which is notably superior to that of the optimal 3-layer InAs structure with $\overline{\eta} = -16.9\%$ and that of the state-of-the-art 10-layer InAs structure with $\overline{\eta} = -15.4\%$.

As shown in \cref{fig:Opt3InAs3Weyl}(b) and \cref{fig:AbsorpEmit3InAs3Weyl}, the use of three InAs layers atop three Weyl semimetal layers enhances the nonreciprocity at wavelengths in both the lower and upper regions of the considered spectrum. Specifically, the top InAs layers improve the absorptivity at wavelength values from 25 to 40 \si{\um}, while the bottom Weyl semimetal layers focus on improving the absorptivity at wavelengths from 5 to 15 \si{\um}.

\cref{table2} lists the optimal parameters for the best 6-layer 3InAs+3Weyl structures obtained from LCB and PI.
While the two acquisition functions yield the same optimal contrast value ($\overline{\eta} = -24.4\%$), they provide two different sets of optimal parameters.
Nevertheless, the carrier concentrations of the InAs layers of these optimal structures still increase linearly from the top to the bottom. Additionally, the thickness of the Weyl semimetal layers reaches the lower bound of its defined domain, indicating that further improvement may be possible with even thinner layers.

\section{Conclusion}

We present an optimization approach combining BO and reparameterization that demonstrates optimal broadband nonreciprocal thermal emitter performance, surpassing state-of-the-art broadband nonreciprocal effects in the infrared range of thermal radiation using doped InAs and Weyl semimetal materials. 
Starting from a less effective structure, the proposed approach incrementally improves the broadband nonreciprocity of the structure by repeatedly reparameterizing the normalized contrast, constructing a probabilistic surrogate model for the reparameterized contrast, maximizing a cheap-to-compute acquisition function formulated from the constructed surrogate model to identify a promising new structure, and updating the surrogate model with the new structure. The optimal structure is the best structure among those recommended by the optimization algorithm upon its termination.
Optimization results indicate that our approach can propose an optimal structure of only three InAs layers that outperforms the current state-of-the-art 10-layer InAs structure \cite{LiuM2023}.
Additionally, the broadband nonreciprocal effect considerably increases when using InAs and Weyl semimetals, which shows the feasibility of combining different nonreciprocal materials for enhanced nonreciprocity. The significant improvements in the designed emitters highlight the role of numerical optimization in advancing practical nonreciprocal thermal emitter development. Our approach can also be adapted to optimize more general structures containing patterns. 
In an ongoing study, we conduct experiments to validate the performance of the optimized structures. We then combine the experimental and numerical observations within the framework of multi-fidelity BO \cite{Huang2006smo,Forrester2007,Kandasamy2017,Do2023,ZhangRD2024mfml} to further enhance our design approach.

\begin{acknowledgement}
The authors acknowledge the funding from the University of Houston through the SEED program and the National Science Foundation under Grant No. CBET-2314210, and the support of the Research Computing Data Core at the University of Houston for assistance with the calculations carried out in this work.
\end{acknowledgement}

\section{Associated content}

\paragraph{Data availability statement}
The data underlying this study are not publicly available as they form part of an ongoing study.
The code for Bayesian optimization is available from the corresponding author upon reasonable request.

\paragraph{Supporting information}
The mathematical foundation of Gaussian process models, the expressions of LCB and PI acquisition functions, the quadrature method for computing the contrast between absorptivity and emissivity at a particular value of design parameters, the details of the reparameterization scheme, and additional results for 5-layer InAs and 8-layer 5InAs+3Weyl structures.

\clearpage
\section{Supporting Information}
\renewcommand{\thefigure}{S\arabic{figure}}
\setcounter{figure}{0}
\renewcommand{\theequation}{S\arabic{equation}}
 \setcounter{page}{1}
\renewcommand{\thepage}{S-\arabic{page}}

\subsection*{Gaussian process}

A Gaussian process (GP) model \cite{Rasmussen2006} is a non-parametric, probabilistic model that enables the prediction of an unknown function of interest $f({\bf x})$ for any ${\bf x} \in \mathbb{R}^d$ and quantifies the uncertainty in that prediction. 
In this work, $f({\bf x})$ represents the normalized contrast function $\overline{\eta}({\bf x})$, see problem (12) in \textbf{Optimization problem}.
The GP model can also be viewed as a distribution over functions, extending the concept of finite-dimensional Gaussians to an infinite-dimensional space. It assumes that the function values
${\bf f} = \left\{ f({\bf x}_i)\right\}_{i=1}^{N}$ evaluated at any finite set ${\bf X} = \left\{ {\bf x}_i\right\}_{i=1}^{N}$ of input variables are distributed according to a multivariate Gaussian. This assumption is encoded in the following GP prior:
\begin{equation}
	f({\bf x}) \sim \mathcal{GP} \left(m({\bf x}),\kappa({\bf x},{\bf x}'|\boldsymbol{\theta})\right),
\end{equation}
where $m({\bf x})$ and $\kappa({\bf x},{\bf x}'|\boldsymbol{\theta})$ are the mean and covariance functions, respectively. The mean function $m({\bf x})$ determines the expected value of $f({\bf x})$ at any location ${\bf x}$ and is often set as $m({\bf x}) = 0$. The covariance function $\kappa({\bf x},{\bf x}'|\boldsymbol{\theta})$ defines the covariance between $f({\bf x})$ and $f({\bf x}')$ and is characterized by a set of parameters $\boldsymbol{\theta}$, known as hyperparameters.
Hereafter, we omit the dependence of $\kappa$ on $\boldsymbol{\theta}$ for clarity.

The covariance function profoundly impacts important properties of a GP model such as continuity and differentiability of its sample paths \cite{Garnett2023}. These properties, in turn, affect the uniqueness and existence of a global solution. A popular choice is the squared exponential (SE) covariance function, which reads
\begin{equation}\label{eqn:s4}
    \kappa({\bf x},{\bf x}') = \sigma_\text{f}^2 \exp \left( -\frac{1}{2}\displaystyle\sum_{k=1}^{d}\frac{(x_k-x_k')^2}{l_j^2} \right),
\end{equation}
where $\sigma_\text{f}$ is the scaling standard deviation and $l_j$ are the characteristic length scales. The hyperparameters for the SE covariance function is $\boldsymbol{\theta} = [\sigma_\text{f},l_1,\dots,l_d]^T$.

We now wish to find a most-likely set for the hyperparameters $\boldsymbol{\theta}$ given the observations ${\bf y} = \left\{ y({\bf x}_i)\right\}_{i=1}^{N}$ for the function values ${\bf f} = \left\{ f({\bf x}_i)\right\}_{i=1}^{N}$. We let $p({\bf f}|{\bf X}) = \mathcal{N} \left( {\bf 0}, {\bf K}({\bf X},{\bf X}) \right)$ represent the zero-mean GP prior from \cref{eqn:s4}, where $p(\cdot)$ represents a probability density function, $\mathcal{N}(\cdot)$ denotes a Gaussian, and the $(i,j)$th entry of ${\bf K}$ is ${\bf K}_{[ij]} = \kappa({\bf x}_i,{\bf x}_j)$ $(i,j=1,\dots,N)$. We further assume that ${\bf y}$ are generated from an observation model 
$p({\bf y}|{\bf f},{\bf X}) = \mathcal{N} \left( {\bf f}, \sigma_\text{n}^2 {\bf I}\right)$ known as the likelihood, where $\sigma_\text{n}^2$ and ${\bf I}$ represent the observation variance and the identity matrix, respectively. Therefore, we can derive the marginal likelihood (for a zero-mean function) $p({\bf y}|{\bf X}) = \int p({\bf f}|{\bf X}) p({\bf y}|{\bf f},{\bf X}) d{\bf f} = \mathcal{N} \left( {\bf 0}, {\bf C}({\bf X},{\bf X}) \right)$ and compute the associated log-marginal likelihood $\log p({\bf y}|{\bf X})$, where ${\bf C}({\bf X},{\bf X}) = {\bf K}({\bf X},{\bf X}) + \sigma_\text{n}^2 {\bf I}$. We then maximize $\log p({\bf y}|{\bf X})$ for a set of most-likely hyperparameters $\boldsymbol{\theta}$ given the observations, which is called the maximum likelihood estimate of the hyperparameters.

Once the hyperparameters have been chosen, the posterior $p\left({\bf f}|{\bf X}, {\bf y}\right)$ can be obtained from Bayes' rule, such that
\begin{equation}\label{eqn:s4}
    p\left({\bf f}|{\bf X}, {\bf y}\right) =
    \frac{p({\bf f}|{\bf X}) p({\bf y}|{\bf f},{\bf X})}{p({\bf y}|{\bf X})}.
\end{equation}
In this case, the posterior $p\left({\bf f}|{\bf X}, {\bf y}\right)$ is a Gaussian because $p({\bf f}|{\bf X})$ and $p({\bf y}|{\bf f},{\bf X})$ are Gaussians. Thus, the posterior is also a GP, which reads \cite{Rasmussen2006}
\begin{equation}
	\widehat{f}({\bf x}) \sim \mathcal{GP} \left(\widehat{m}({\bf x}),\widehat{\kappa}({\bf x},{\bf x}')\right),
\end{equation}
where the conditional mean $\widehat{m}({\bf x})= {\bf K}^T_{\bf{x}} {\bf C}^{-1} {\bf y}$ and the condictional covariance $\widehat{\kappa}({\bf x}, {\bf x}')=\kappa({\bf x},{\bf x}') -{\bf K}^T_{{\bf x}} {\bf C}^{-1} {\bf K}_{{\bf x}'}$ with ${\bf K}_{{\bf x}} = \left[\kappa({\bf x},{\bf x}_1),\dots,\kappa({\bf x},{\bf x}_N)\right]^T$ and ${\bf K}_{{\bf x}'} = \left[\kappa({\bf x}',{\bf x}_1),\dots,\kappa({\bf x}',{\bf x}_N)\right]^T$.
The prediction $f_\star$ at a new point ${\bf x}_\star$
can be calculated from the GP posterior as $p(f_\star|\widehat{f},{\bf x}_\star) = \mathcal{N} \left( \widehat{\mu}({\bf x}_\star), \widehat{\sigma}^2({\bf x}_\star) \right)$, where $\widehat{\mu}({\bf x}_\star) = \widehat{m}({\bf x}_\star)$ and $\widehat{\sigma}^2({\bf x}_\star) = \widehat{\kappa}({\bf x}_\star,{\bf x}_\star)$ are the predictive mean and variance, respectively.

\subsection*{Lower confidence bound and probability of improvement}

In this section, we describe the lower confidence bound (LCB) and probability of improvement (PI) acquisition functions BO uses for optimizing the normalized contrast $\overline{\eta} ({\bf x})$ in \textbf{Initialization}.

The LCB acquisition function balances greedy optimization
introduced by minimizing the predictive mean function $\widehat{\mu}({\bf x})$ (i.e., maximizing $-\widehat{\mu}({\bf x})$) with uncertainty reduction employed by maximizing the predictive standard deviation function $\widehat{\sigma}({\bf x})$.
It is a variant of the GP-lower confidence bound \cite{Srinivas2010}, which has been proposed for maximization problems. The LCB acquisition function reads
\begin{equation}
	q({\bf x}) = -\left(\widehat{\mu}({\bf x}) - \beta \widehat{\sigma}({\bf x})\right),
\end{equation}
where $\beta \geq 0$ is a tuning parameter trading off between greedy optimization and uncertainty reduction.

The PI acquisition function~\citep{Kushner1964} measures the probability that the one-step lookahead observation of the objective function $f({\bf x})$ recommended by BO is better than the best value $f_{\min}$ it has observed so far.
This is equivalent to measuring the chance of having a solution improvement in the next iteration of BO.
Conditioning PI on the current GP posterior $\widehat{f}({\bf x})$, it can be analytically written as
\begin{equation}
	q({\bf x}) = \mathbb{P}\left[f({\bf x})<f_{\min}\right|\widehat{f}({\bf x})]=
    \Phi\left(\frac{f_{\min} - \widehat{\mu}({\bf x})}{\widehat{\sigma}({\bf x})}\right),
\end{equation}
where $\Phi(\cdot)$ represents the standard normal cumulative distribution function. 

\subsection*{Quadrature for computing contrast values}

We describe here the numerical quadrature method used to compute the contrast $\eta({\bf p})$
for a specific value of design parameters ${\bf p}$.
Recall that
\begin{equation} \label{eq:totalcontrast} 
\begin{aligned}
    \eta({\bf p}) & = \int_{\lambda_\text{L}}^{\lambda_\text{U}} \int_{0}^{\pi/2} \eta({\bf p},\lambda,\theta) \sin{\theta} \cos{\theta} \diff\theta \diff\lambda\\
    & = \int_{\lambda_\text{L}}^{\lambda_\text{U}} \int_{0}^{\pi/2}
    \left(\alpha({\bf p},\lambda,\theta) - \varepsilon({\bf p},\lambda,\theta)\right)
    \sin{\theta} \cos{\theta} \diff\theta \diff\lambda,
\end{aligned}
\end{equation} 
where $\alpha({\bf p},\lambda,\theta)$ and $\varepsilon({\bf p},\lambda,\theta)$ represent the local functions for the absorptivity and emissivity. Hereafter, we omit the dependence of $\eta$, $\alpha$, and $\varepsilon$ on ${\bf p}$ for clarity.

Computing the integrals in \cref{eq:totalcontrast} requires evaluations of $\alpha$ and $\varepsilon$
at many different $(\lambda, \theta)$ locations, which is costly.
To improve computational efficiency and accuracy, we apply a numerical quadrature scheme for the integrals.
A quadrature scheme approximates an integral $I = \int_a^b f(x) dx$ with a numerical formula $I_n = \sum_{k=1}^n w_k f(x_k)$,
where $x_1, \dots, x_n$ are the nodes at which the integrand is evaluated
and $w_1, \dots, w_n$ are the quadrature weights.

First we separate the integrals of $\alpha(\lambda,\theta)$ and $\varepsilon(\lambda,\theta)$,
because these two functions are less fluctuating than $\eta(\lambda,\theta)$ and thus easier to integrate.
Our quadrature method applies identically to both $\alpha$ and $\varepsilon$,
so we only present the calculation for $\alpha$ below.
Starting with the inner integral of \cref{eq:totalcontrast} with respect to the angle of incidence $\theta$,
we define
\begin{equation}
    \alpha(\lambda) = \int_{0}^{\pi/2} \alpha(\lambda,\theta) \sin{\theta} \cos{\theta} \diff\theta.
\end{equation} 
To approximate this integral, we use Gauss quadrature with $n_\theta$ nodes and the corresponding weights.
For the details of Gauss quadrature, see Chapter 19 in this textbook\cite{Trefethen2019}.

To simplify the computation of the outer integral, we transform wavelength $\lambda$ to wavenumber $\nu$.
This is because $\alpha(\nu)$ and $\eta(\nu)$ are smoother than $\alpha(\lambda)$ and $\eta(\lambda)$, respectively.
Define
\begin{equation}
    \alpha_\lambda  = \int_{\lambda_\text{L}}^{\lambda_\text{U}}  \alpha(\lambda) \diff\lambda.
\end{equation} 
Let $\lambda_1 = s_\lambda \lambda$ and $\nu_1 = \lambda_1^{-1}$,
where the scaling factor $s_\lambda = 10^5$ for $\lambda$ in meters.
This transformation maps the wavelength $\lambda \in [\lambda_\text{L},\lambda_\text{U}]$ $\text{m}$
to $\lambda_1 \in [\lambda_{1\text{L}},\lambda_{1\text{U}}] \times 10^{-5}\text{m}$,
and subsequently to the wavenumber $\nu_1 \in [\nu_{1\text{L}},\nu_{1\text{U}}]$.
Thus, we have
\begin{equation}
    \alpha_\lambda = \int_{\lambda_\text{L}}^{\lambda_\text{U}} \alpha(\lambda) \diff\lambda = s_\lambda^{-1} \alpha_{\lambda_1},
\end{equation}     
where
\begin{equation} \label{eq:totalabsorp} 
    \alpha_{\lambda_1} = \int_{\lambda_{1\text{L}}}^{\lambda_{1\text{U}}} \alpha(\lambda_1) \diff\lambda_1 = \int_{\nu_{1\text{L}}}^{\nu_{1\text{U}}} \alpha(g(\nu_1)) g'(\nu_1) \diff\nu_1.
\end{equation}
Here, $g(\nu_1) = 1/\nu_1$ and $g'(\nu_1) = -1/\nu_1^2$.

We compute the integral in \cref{eq:totalabsorp} with a transformed Gauss quadrature \cite{Hale2008}.
It starts with a linear change of variables from $\nu_1 \in [\nu_{1\text{L}},\nu_{1\text{U}}]$ to $x \in [-1,1]$.
This allows us to utilize the following relationship
between the integrals of a function $k(s)$ over $[-1,1]$ and over a strip map $l(x)$ with $x \in [-1,1]$ to compute $\alpha_{\lambda_1}$:
\begin{equation}
    I(k) = \int_{-1}^{1} k(s) ds = \int_{-1}^{1} k(l(x)) l'(x) dx, 
\end{equation}     
where $k$ and $l$ play the same roles as $\alpha$ and $g$ in \cref{eq:totalabsorp}, respectively. 
The transformed integral is computed using Gauss quadrature with
$n_s$ nodes and the corresponding weights.

All optimization results presented in this study are generated by $n_\theta = 24$ and $n_s = 140$. The significantly higher value of $n_s$ compared to $n_\theta$ is due to the substantial fluctuations in contrast along the wavelength direction.
Based on numerical results, this quadrature scheme achieves a relative error of $\mathcal{O}(10^{-5})$.

\subsection*{Details of the reparameterization scheme}

This section details the reparameterization strategy we propose for optimizing the nonreciprocal emitters presented in this work. 
In \textbf{Optimization problem}, we transform the design parameters governing the carrier concentrations ${\bf N} = \left[ n_{e,1},\dots,n_{e,n_1} \right]^T$ of $n_1$ InAs layers and Fermi levels ${\bf E} = \left[ E_{F,1},\dots,E_{F,n_2} \right]^T$ of $n_2$ Weyl layers into optimization parameters ${\bf v}$ and ${\bf w}$ using two linear maps ${\bf D}$ and ${\bf F}$, respectively. Specifically, we have (see Figure 2)
\begin{equation}
\begin{aligned}
    \mathbf{N} & = \mathbf{D}\mathbf{v} = \sum_{i=0}^{p_1-1}v_i\mathbf{D}_{[i]}, \\
    {\bf E} & = \mathbf{F}\mathbf{w} = \sum_{j=0}^{p_2-1}w_j\mathbf{F}_{[j]},
\end{aligned}
\end{equation}
where $\mathbf{N} \in \mathbb{R}^{n_1}$, ${\bf E} \in \mathbb{R}^{n_2}$, $\mathbf{D} \in \mathbb{R}^{n_1 \times p_1}$, $\mathbf{v} \in \mathbb{R}^{p_1}$, $\mathbf{F} \in \mathbb{R}^{n_2 \times p_2}$, $\mathbf{w} \in \mathbb{R}^{p_2}$, $\mathbf{D}_{[i]} \in \mathbb{R}^{n_1}$ is the $(i+1)$th column of $\mathbf{D}$ $(i=0,\dots,p_1-1)$, and $\mathbf{F}_{[j]} \in \mathbb{R}^{n_2}$ is the $(j+1)$th column of $\mathbf{F}$ $(j=0,\dots,p_2-1)$. To increase the efficiency of BO when optimizing structures of a large number of layers, we set $p_1 < n_1$ and $p_2 < n_2$.

To further control the profiles of the InAs carrier concentrations and the Weyl Fermi levels inside the multilayer structure, we define the $k$th entry of vector $\mathbf{D}_{[i]}$ and the $l$th entry of vector $\mathbf{F}_{[j]}$ using layer-wise integration of Chebyshev polynomials of the first kind, as
\begin{equation}
\begin{aligned}
    \mathbf{D}_{[ik]} &= \int_{u_{k-1}}^{u_{k}}T_i(u) du, \quad i=0,\dots,p_1-1, \ \ k = 1,\dots,n_1, \\
    \mathbf{F}_{[jl]} &= \int_{u_{l-1}}^{u_{l}}T_l(u) du, \quad j=0,\dots,p_2-1, \ \ l = 1,\dots,n_2, 
\end{aligned}
\end{equation}
where $T_i(u)$ represents the $i$th order Chebyshev polynomial of the first kind evaluated at any $u \in [-1,1]$, $u_{k} = 2k/n_1-1$ for $k=1,\dots,n_1$, and $u_{l} = 2l/n_2-1$ for $l=1,\dots,n_2$. The first three Chebyshev polynomials of the first kind are $T_0  = 1$, 
$T_1(u)  = u$, and $T_2(u) = 2u^2 -1$.
Higher order polynomials can be obtained from the recurrence relation $T_{i+1}(u) = 2 u T_i(u) - T_{i-1}(u)$ for $i \geq 2$.

\begin{figure*}[]
	\centering
	\includegraphics[scale=0.53]{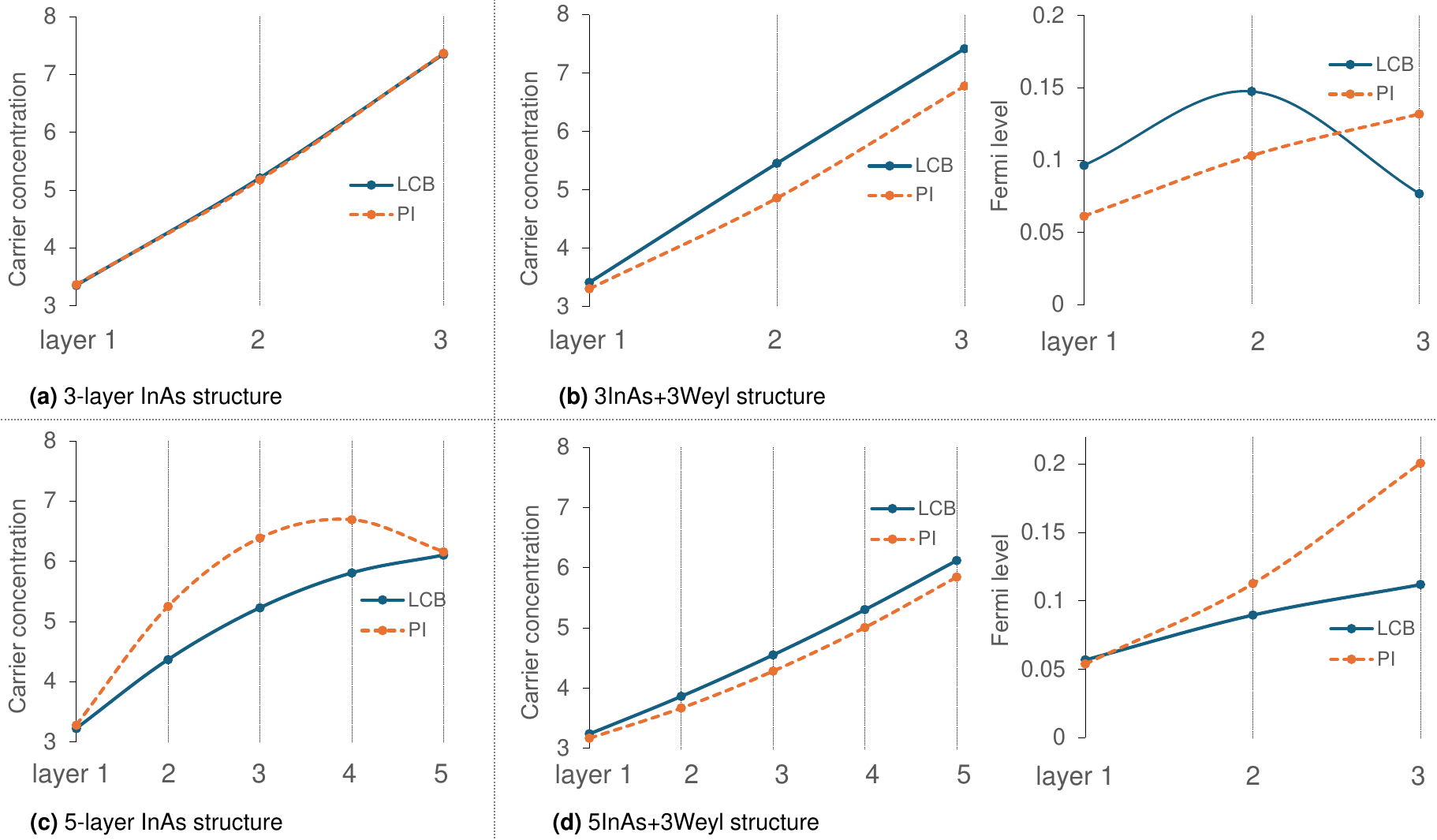}
	\caption{Profiles of InAs carrier concentrations and Weyl Fermi levels inside the optimal structures by LCB and PI.}
    \label{fig:paradistribution}
\end{figure*}

We set $p_1 = 3$ for optimizing the 3-layer InAs structure, and $p_1 = p_2 = 3$ for the 6-layer 3InAs+3Weyl structure presented in \textbf{Results and discussion}. For the 5-layer InAs and 8-layer 5InAs+3Weyl structures presented in \textbf{Additional results} below, we also set $p_1 = 3$ and $p_1 = p_2 = 3$, respectively.
These settings allow the InAs carrier concentrations and Weyl Fermi levels over the layers to conform to a polynomial of up to second order, see \cref{fig:paradistribution}.

\subsection*{Additional results}

\begin{figure*}[t]
	\centering
	\includegraphics[scale=0.9]{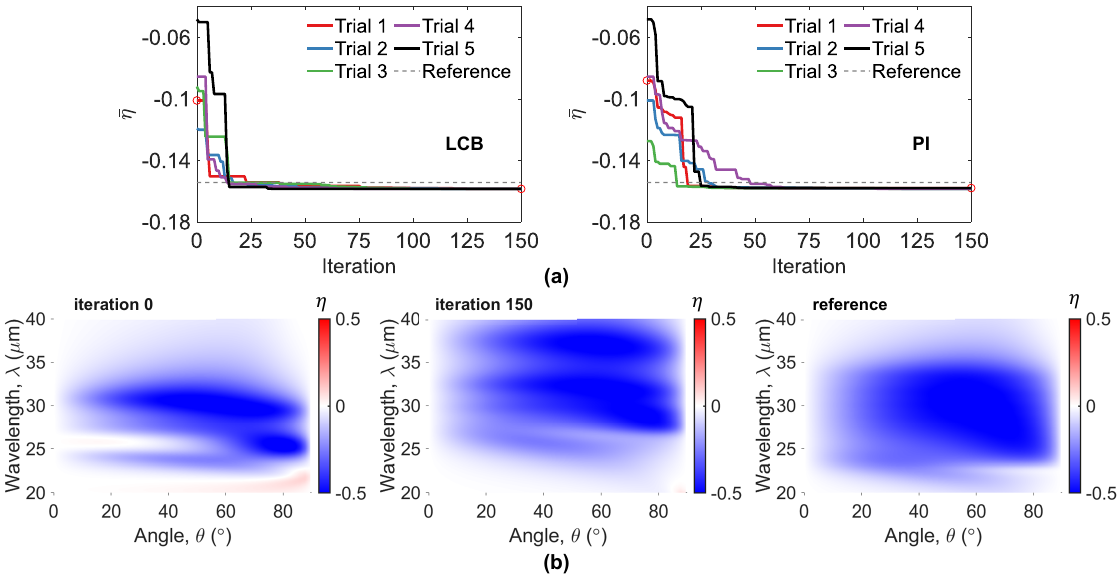}
	\caption{Optimization results for 5-layer InAs structure in comparison with the state-of-the-art 10-layer InAs structure (reference) \cite{LiuM2023}. (a) Optimization histories from LCB and PI. (b) Comparison of contrast values of the initial and final structures from the first and last iterations of the first LCB trial, and the state-of-the-art 10-layer InAs structure.}
    \label{fig:Opt5InAs}
\end{figure*}

\begin{figure*}[t]
	\centering
	\includegraphics[scale=1]{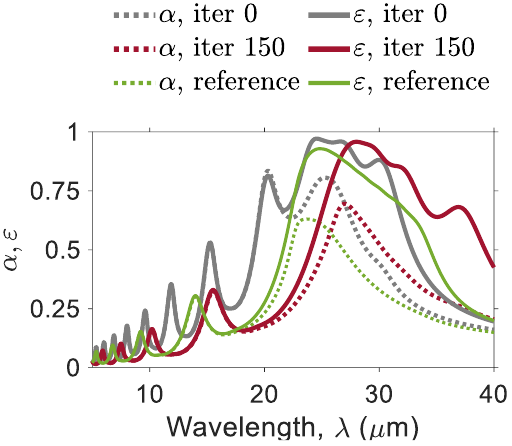}
	\caption{Absorptivity ($\alpha$) and emissivity ($\varepsilon$) of the initial and final 5-layer InAs structures from the first and last iterations of the first LCB trial, and the state-of-the-art 10-layer InAs structure (reference) \cite{LiuM2023}.}
    \label{fig:AbsorpEmit5InAs}
\end{figure*}

\begin{figure*}[t]
	\centering
	\includegraphics[scale=0.9]{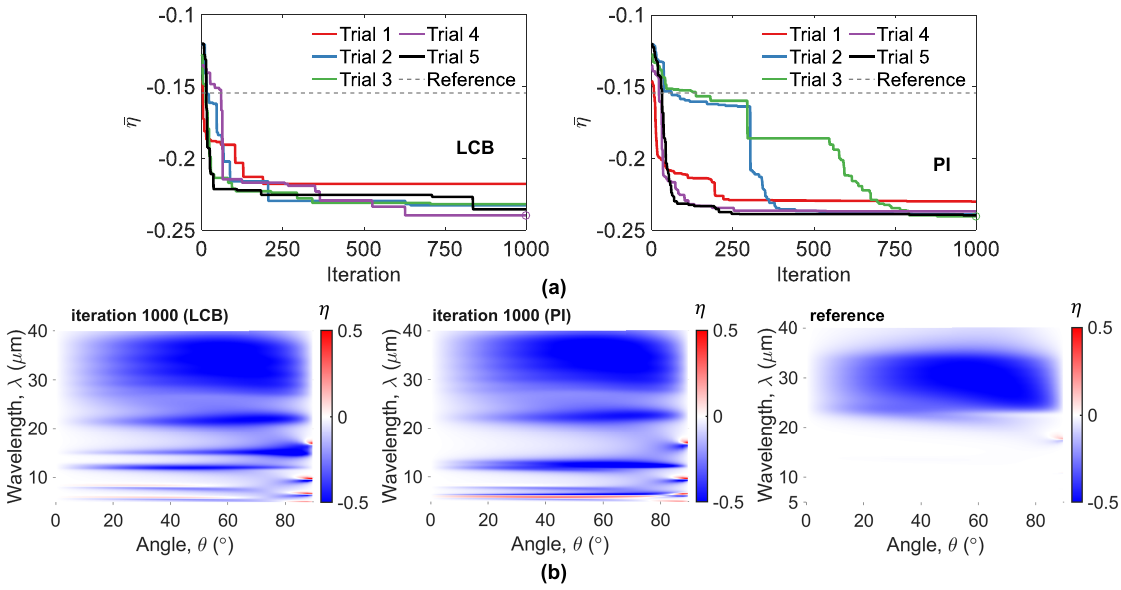}
	\caption{Optimization results for 8-layer 5InAs+3Weyl structure in comparison with the state-of-the-art 10-layer InAs structure (reference) \cite{LiuM2023}. (a) Optimization histories from LCB and PI. (b) Comparison of contrast values of the best structures from LCB and PI, and the state-of-the-art 10-layer InAs structure.}
    \label{fig:Opt5InAs3Weyl}
\end{figure*}

\begin{figure*}[t]
	\centering
	\includegraphics[scale=1]{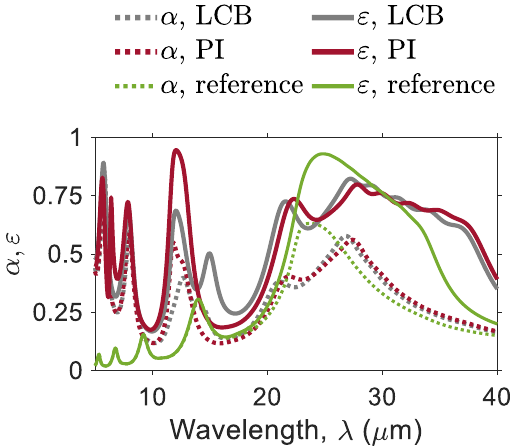}
	\caption{Absorptivity ($\alpha$) and emissivity ($\varepsilon$) of the best 8-layer 5InAs+3Weyl structures from LCB and PI, and the state-of-the-art 10-layer InAs structure (reference) \cite{LiuM2023}.}
    \label{fig:AbsorpEmit5InAs3Weyl}
\end{figure*}

This section provides optimization results for 5-layer InAs and 8-layer 5InAs+3Weyl structures. 
\Cref{fig:Opt5InAs} shows the results for the 5-layer InAs structures obtained from five different trials of LCB and those of PI.
\Cref{fig:AbsorpEmit5InAs} plots the absorptivity and emissivity values over the considered range of incidence angles for the initial and final 5-layer InAs structures from the first and last iterations of the first LCB trial, as well as those for the state-of-the-art 10-layer InAs structure \cite{LiuM2023}.
\Cref{fig:Opt5InAs3Weyl} shows the optimization results for the 8-layer 5InAs+3Weyl structures.
The absorptivity and emissivity values over the considered range of incidence angles for the best 8-layer 5InAs+3Weyl structures from LCB and PI are provided in \cref{fig:AbsorpEmit5InAs3Weyl}.

\clearpage
\bibliography{BOPhotonics}

\end{document}